\documentclass[showpacs,pra,aps,superscriptaddress,floatfix,twocolumn]{revtex4}
\usepackage{amsmath}
\usepackage{amssymb,mathrsfs}
\usepackage{graphicx}
\usepackage[dvipsnames]{xcolor}

\usepackage[utf8]{inputenc}
\usepackage[english]{babel}

\newcommand{\prt}{\partial}

\newcommand{\la}{\lambda}

\newcommand{\ga}{\gamma}

\begin{document}
\title{Theory of dispersive shock waves induced by the Raman effect in optical fibers }

\author{D. V. Shaykin}
\email{shaykin.dv@phystech.edu}
\affiliation{Russian University of Transport (RUT-MIIT), Obrazcova st. 9, Moscow, 127994, Russia}
\affiliation{ Moscow Institute of Physics and Technology, Institutsky Lane 9, Dolgoprudny, Moscow Region 141700, Russia}
\affiliation{Skolkovo Institute of Science and Technology, Skolkovo, Moscow, 143026, Russia}

\author{A. M. Kamchatnov}
\email{kamch@isan.troitsk.ru}
\affiliation{Institute of Spectroscopy Russian Academy of Sciences, Troitsk,
Moscow, 108840, Russia}
\affiliation{Skolkovo Institute of Science and Technology, Skolkovo, Moscow, 143026, Russia}
\affiliation{ Higher School of Economics, Physical Department, 20 Myasnitskaya ulica, Moscow 101000, Russia}

\begin{abstract}
We develop the theory of dispersive shock waves in optical fibers for the case of long-distance
propagation of optical pulses, when the small Raman effect stabilizes the profile of the shock.
The Whitham modulation equations are derived as the basis for the Gurevich-Pitaevskii approach
to the analytical theory of such shocks. We show that the wave variables at both sides of the
shock are related by the analogue of the Rankine-Hugoniot condition that follows from the
conservation laws of the Whitham equations. Solutions of the Whitham equations yield the
profiles of the wave variables that agree very well with the exact numerical solution of the
generalized nonlinear Schr\"{o}dinger equation for propagation of optical pulses.
\end{abstract}


\maketitle

\section{Introduction}

Nonlinear wave structures called dispersive shock waves (DSWs) have been observed in a number of different
physical media, from water waves to Bose-Einstein condensates (see, e.g., review articles \cite{eh-16,kamch-21}
and references therein). Generally speaking, they can be represented as a lengthy oscillatory nonlinear wave
structure that degenerates at one of its edges to a train of solitons and at the other edge to a small-amplitude wavy 
tail.  If such a DSW is formed as a result of a wave breaking of a large-scale wave pulse, so that at the initial 
stage  of evolution dispersion effects dominate over dissipative ones, then this DSW expands with time with an
increasing number of oscillations in it. Typically, the wavelength in a DSW is much smaller than its whole size;
therefore, this DSW can be represented as a modulated nonlinear periodic wave with parameters (amplitude of
oscillations, wavelength, etc.) slowly changing with space and time. However, even small dissipation becomes
crucially important at the later stage of evolution of DSWs, when their slow dynamics due to modulation become
comparable with slow dynamics due to small dissipation. As a result, a DSW stops its expansion and tends to a
stationary wave structure whose total length is proportional to the inverse of the small dissipation parameter.
In both cases, the modulation is small, and this fact was used by Gurevich and Pitaevskii \cite{gp-73} in their
approach to the theory of DSWs based on Whitham's theory of modulations of periodic solutions of nonlinear wave
equations \cite{whitham-65,whitham-74}, in particular, of the Korteweg-de Vries (KdV) equation. This approach
turned out very successful in the theoretical description of DSWs described by the KdV equation both in
time-dependent \cite{gle-90,wright-93,tian-93} and stationary situations with account of small dissipation
\cite{gp-87,akn-87,mg-95,kamch-16}.

DSWs in optical fibers were first observed long ago \cite{tsj-85,rg-89}, but their theoretical description was a
difficult problem since its solution needed application of the quite involved inverse scattering transform method,
discovered in Refs.~\cite{ggkm-67,lax-68,zs-71,zs-73}, to the nonlinear Schr\"{o}dinger (NLS) equation, which
describes propagation of light pulses in fibers. Whitham modulation equations for the NLS case without any
perturbations were obtained in Refs.~\cite{fl-86,pavlov-87}, and their solution for the important problem of
evolution of an initial discontinuity was found in Refs.~\cite{ak-87,eggk-95}. This theory was confirmed in
the optical experiment \cite{xckt-17} with initial pulses having a specially engineered sharp ``discontinuity''.
More general forms of DSWs were studied theoretically, e.g., in Ref.~\cite{ikp-19}, and experimentally in
Ref.~\cite{bienaime-21}. However, optical DSWs with dissipation have not been studied much so far, because in
the optical case standard forms of dissipation also affect a smooth part of a pulse rather than only the
strongly oscillatory region. A quite specific situation of the formation of DSWs by the flow of polariton
fluid past an obstacle when dissipation was compensated by pumping was discussed in Ref.~\cite{lpk-12}.

As was found in Refs.~\cite{dianov-85,mm-86}, the induced Raman scattering in fibers can play the role
of pumping or dissipation. In case of normal group velocity dispersion, formation of dark solitons at sharp
edges of a pulse was observed in Ref.~\cite{weiner-89}. Propagation of pulses in fibers with account of
induced Raman scattering is described by the equation \cite{stolen-89,kivshar-90} (see also \cite{ka-03})
\begin{equation}\label{eq1}
  i\psi_x+\frac12\psi_{tt}-|\psi|^2\psi=-\ga\psi(|\psi|^2)_t,
\end{equation}
written here in standard non-dimensional form for the case of normal dispersion. Here $\psi$ denotes the
strength of an electromagnetic wave in a fiber, $x$ is a coordinate along the fiber, $t$ is the normalized time,
and $\ga$ is a small parameter that measures the delay in the Raman response function. Thus, the right-hand side
of Eq.~(\ref{eq1}) can be considered as a small perturbation of the standard NLS equation. If small-amplitude
waves propagating along a large-scale uniform background are considered, then this equation can be reduced to the
KdV equation with Burgers dissipation, and the formation of DSWs was predicted in Refs.~\cite{kivshar-90,km-93}.
The aim of this paper is to develop the theory of DSWs described by Eq.~(\ref{eq1}) without any restrictions on
the amplitude of waves in the framework of the Gurevich-Pitaevskii approach. In Section~\ref{sec2} we describe the
periodic solutions of the NLS equation, in Section~\ref{sec3} we derive the Whitham equations for modulations of
these solutions with account of the Raman term, and in Section~\ref{sec4} we find stationary solutions of the
Whitham equations. Our analytical theory agrees very well with numerical solutions of Eq.~(\ref{eq1}).

\section{Periodic solutions of the NLS equation}\label{sec2}

If we neglect small Raman effect, then Eq.~(\ref{eq1}) with $\ga=0$ reduces to the standard NLS equation
\begin{equation}\label{eq2}
  i\psi_x+\frac12\psi_{tt}-|\psi|^2\psi=0
\end{equation}
with exchanged roles of the space ($x$) and time ($t$) variables. The Madelung transformation
\begin{equation}\label{eq3}
    \psi(x,t) = \sqrt{\rho(x,t)}\exp{\left({ i}\int^t u(x,t')dt'\right)}
\end{equation}
casts it to the hydrodynamic-like form
\begin{equation} \label{eq4}
\begin{split}
    & \rho_x+(u\rho)_t=0, \\
    & u_x+uu_t+\rho_t+\left(\frac{\rho^2_t}{8\rho^2}-\frac{\rho_{tt}}{4\rho}\right)_t=0.
\end{split}
\end{equation}
where $\rho=\rho(x,t)$ is the intensity of light and $u=u(x,t)$ is the chirp. If the light pulse is
smooth enough, then we can neglect the terms with 3rd order derivatives that describe dispersion effects
and obtain hydrodynamic equations of the dispersionless approximation
\begin{equation} \label{eq4b}
\begin{split}
     \rho_x+(u\rho)_t=0, \quad
     u_x+uu_t+\rho_t=0.
\end{split}
\end{equation}
They have the form of the so-called ``shallow water'' equations with exchanged $x$ and $t$ variables
(see, e.g., Refs.~\cite{whitham-74,kamch-2024}) and take especially simple diagonal form
\begin{equation}\label{eq4c}
  \frac{\prt\la_+}{\prt x}+\frac{1}{v_+}\frac{\prt\la_+}{\prt t}=0,\quad
  \frac{\prt\la_-}{\prt x}+\frac{1}{v_-}\frac{\prt\la_-}{\prt t}=0
\end{equation}
in terms of the Riemann invariants
\begin{equation}\label{eq4d}
  \la_{\pm}=\frac{u}{2}\pm\sqrt{\rho},
\end{equation}
where
\begin{equation}\label{eq4e}
  \frac{1}{v_{\pm}}=u\pm\sqrt{\rho}.
\end{equation}
We assume that at both sides of the DSW far enough from its leading front, the distributions of
$\rho$ and $u$ are uniform, so the Riemann invariants are constant,
\begin{equation}\label{eq4f}
  \la\to\left\{
  \begin{array}{l}
   \la_{\pm}^L=u^L/2\pm\sqrt{\rho^L}\quad\text{for}\quad t\to-\infty,\\
   \la_{\pm}^R=u^R/2\pm\sqrt{\rho^R}\quad\text{for}\quad t\to+\infty.
  \end{array}
  \right.
\end{equation}
These two flows of ``fluid of light'' are joined by a single DSW for a proper choice of
the boundary conditions $\rho^L,u^L,\rho^R,u^R$.

In Gurevich-Pitaevskii
approach, a DSW is represented as a modulated periodic solution of Eqs.~(\ref{eq4}) and we take this
solution in the form most convenient for the modulation theory (see, e.g., \cite{kamch-2000,kamch-2024})
\begin{equation}\label{eq5}
\begin{split}
    \rho=&\frac{1}{4}(\lambda_4-\lambda_3-\lambda_2+\lambda_1)^2+(\lambda_4-\lambda_3)(\lambda_2-\lambda_1)\\
    &\times
    \mathrm{sn}^2(\sqrt{(\lambda_4-\lambda_2)(\lambda_3-\lambda_1)}\theta,m), \\
    u=&V-\frac{j}{\rho},
    \end{split}
    \end{equation}
where
\begin{equation}\label{eq6}
\begin{split}
    & \theta=t-\frac{x}{V},\qquad \frac{1}{V}=\frac{1}{2}\sum_{i=1}^{4}\lambda_i,\\
    & m=\frac{(\lambda_2-\lambda_1)(\lambda_4-\lambda_3)}{(\lambda_4-\lambda_2)(\lambda_3-\lambda_1)},\quad 0\leq m\leq1; \\
    & j=\frac{1}{8}(-\lambda_1-\lambda_2+\lambda_3+\lambda_4)\times\\
    &\times(-\lambda_1+\lambda_2-\lambda_3+\lambda_4)
    (\lambda_1-\lambda_2-\lambda_3+\lambda_4).
\end{split}
\end{equation}
The parameters $\la_i,i=1,2,3,4,$ are ordered according to inequalities
\begin{equation}\label{eq7}
    \lambda_1\leq\lambda_2\leq\lambda_3\leq\lambda_4;
\end{equation}
they are constant in a non-modulated wave and change slowly in a slightly modulated one.
As is clear from Eqs.~(\ref{eq5}) and (\ref{eq6}), the phase velocity $V$ and the amplitude
\begin{equation}\label{eq7b}
  a=(\lambda_4-\lambda_3)(\lambda_2-\lambda_1)
\end{equation}
of the nonlinear wave are expressed in terms of these parameters. The Jacobi elliptic function
$\mathrm{sn}$ is periodic, so we get for the period the expression
\begin{equation} \label{eq8}
    T=\frac{2K(m)}{\sqrt{(\lambda_1-\lambda_3)(\lambda_2-\lambda_4)}},
\end{equation}
where $K(m)$ is the complete elliptic integral of the first kind. The soliton solution
corresponds to the limiting case with $\la_2=\la_3$, so it can be written in the form
\begin{equation}\label{eq8b}\begin{split}
  \rho=&\frac14(\la_4-\la_1)^2\\
  &-\frac{(\la_4-\la_2)(\la_2-\la_1)}{\cosh^2[\sqrt{(\la_4-\la_2)(\la_2-\la_1)}(t-x/v_s)},
  \end{split}
\end{equation}
and its velocity is given by
\begin{equation}\label{eq8c}
  \frac{1}{v_s}=\la_2+\frac12(\la_1+\la_4).
\end{equation}
Although these formulas can be
obtained by direct finding the traveling wave solutions of Eqs.~(\ref{eq4}), we will need for
derivation of the modulation equations some more subtle results following from the complete
integrability of the NLS equation (\ref{eq2}).

First of all, we notice that Eq.~(\ref{eq2}) can be written as a compatibility condition
$(\phi_{xx})_t=(\phi_t)_{xx}$ of two linear differential equations
\begin{equation} \label{eq9}
\begin{split}
    \phi_{tt} & = \mathcal{A}\phi, \\
    \phi_x & = -\frac12 \mathcal{B}_t\phi+\mathcal{B}\phi_t,
\end{split}
\end{equation}
where $\mathcal{A}$ and $\mathcal{B}$ depend on the field variables $\psi,\psi^*$ and the
spectral parameter $\la$,
\begin{equation} \label{eq10}
\begin{split}
    \mathcal{A} & = -\lambda^2 + i\lambda\frac{\psi_t}{\psi} + |\psi|^2 -
    \frac{1}{2}\frac{\psi_{tt}}{\psi} + \frac{3}{4}\frac{\psi_t^2}{\psi^2}, \\
    \mathcal{B} & = -\lambda + \frac{i}{2}\frac{\psi_t}{\psi},
\end{split}
\end{equation}
so the compatibility condition is fulfilled for any value of $\la$. The first Eq.~(\ref{eq9})
has two basis solutions $\phi_+$ and $\phi_-$, and it is easy to check that their product
$g=\phi_+\phi_-$ satisfies the equation
\begin{equation}\nonumber
  g_{ttt}-2\mathcal{A}_tg-4\mathcal{A}g_t=0,
\end{equation}
which can be easily integrated once to give
\begin{equation} \label{eq11}
\begin{split}
    \frac{1}{2}gg_{tt}-\frac{1}{4}g_t^2
    -\mathcal{A}g^2= P(\lambda).
\end{split}
\end{equation}
Periodic solutions of Eq.~(\ref{eq2}) are distinguished by the condition that the integration
constant $P(\la)$ must be a polynomial in the spectral parameter $\la$. The solution (\ref{eq5})
corresponds to the 4th degree polynomial
\begin{equation} \label{eq12}
    P(\lambda)=\prod_{i=1}^{4}(\lambda-\lambda_i)=\lambda^4-s_1\lambda^3+s_2\lambda^2-s_3\lambda+s_4,
\end{equation}
where
\begin{equation} \label{eq13}
\begin{split}
    & s_1=\sum_i\lambda_i,\quad s_2=\sum_{i<j}\lambda_i\lambda_j,
    \quad s_3=\sum_{i<j<k}\lambda_i\lambda_j\lambda_k, \\
    & s_4=\lambda_1\lambda_2\lambda_3\lambda_4.
\end{split}
\end{equation}
Then $g$ is the 1st degree polynomial
\begin{equation}\label{eq14}
  g=\la-\mu(\theta),\quad \theta=t-\frac{x}{V},\quad \frac{1}{V}=\frac{s_1}{2},
\end{equation}
which satisfies the equation
\begin{equation}\label{eq15}
  \frac{d\mu}{d\theta}=2\sqrt{-P(\mu)}.
\end{equation}
The variable $\mu$ is complex and with change of $\theta$ it moves in the complex plane around
a locus defined by the formula
\begin{equation} \label{15}
\begin{split}
    \mu(\rho)  = \frac{s_1}{4}+\frac{-j+i\sqrt{\mathcal{R}(\rho)}}{2\rho},
\end{split}
\end{equation}
where $\rho$ is given by Eq.~(\ref{eq5}). It satisfies the equation
\begin{equation}\label{eq17}
\begin{split}
 &\frac{d\rho}{d\theta} = 2\sqrt{\mathcal{R}(\rho)}, \\
  &  \mathcal{R}(\rho)  = (\rho-\nu_1)(\rho-\nu_2)(\rho-\nu_3),
    \end{split}
\end{equation}
where $\nu_i$ are zeroes of the 3rd degree polynomial $\mathcal{R}(\rho)$ related to
$\lambda_i$ by the formulas
\begin{equation} \label{eq18}
\begin{split}
    \nu_1=\frac14 (\lambda_1-\lambda_2-\lambda_3+\lambda_4)^2, \\
    \nu_2=\frac14 (\lambda_1-\lambda_2+\lambda_3-\lambda_4)^2, \\
    \nu_3=\frac14 (\lambda_1+\lambda_2-\lambda_3-\lambda_4)^2.
\end{split}
\end{equation}
At last, we will also need the formula
\begin{equation}\label{eq19}
  \frac{i\psi_t}{2\psi}=\mu-\frac{s_1}{2},
\end{equation}
relating the field variable $\psi$ with $\mu$.

Now we can turn to derivation of the Whitham modulation equations.

\section{Whitham modulation equations}\label{sec3}

Direct Whitham's method of derivation of modulation equations is not effective in case of
Eq.~(\ref{eq1}), so we turn to the general method developed in Ref.~\cite{kamch-04}
in framework of the Ablowitz-Kaup-Newell-Segur (AKNS) scheme \cite{akns-74}. Earlier,
it was applied to the Kaup-Boussinesq-Burgers equation for shallow water in Ref.~\cite{egk-05}
and to flow of polariton condensate past an obstacle in Ref.~\cite{lpk-12} (see also
Ref.~\cite{kamch-2024}). Here we will apply it to the NLS equation (\ref{eq1}) with
account of induced Raman scattering in light fibers.

In a modulated DSW the parameters $\la_1, i=1,2,3,4,$ become slow functions of $x$ and $t$,
so their evolution obeys the Whitham equations which for completely integrable equations
of NLS type can be written in the form \cite{kamch-04}
\begin{equation}\label{eq20}
\begin{split}
  &\frac{\prt\la_i}{\prt x}+\frac1{v_i}\frac{\prt\la_i}{\prt t}=
  \frac{1}{\langle1/{{g}}\rangle\prod_{j\neq i}(\la_i-\la_j)}\\
  &\times\sum_{m=1}^2\sum_{l=0}^{A_m}\Big\langle\Big\{\frac{\prt\mathcal{A}}
  {\prt \psi_m^{(l)}}\frac{\prt^lR_m}
  {\prt t^l}\Big\}{{g}}\Big\rangle, \quad i=1,\ldots,4,
  \end{split}
\end{equation}
where angle brackets denote averaging over the period of the wave.
In our case we have two wave variables $\psi_1=\psi$ and $\psi_2=\psi^*$ whose evolution
is governed by the equations
\begin{equation}\label{eq21}
  \begin{split}
  \psi_x&=\frac{i}{2}\psi_{tt}-i|\psi|^2\psi+i\ga\psi(|\psi|^2)_t,\\
  \psi^*_x&=-\frac{i}{2}\psi^*_{tt}+i|\psi|^2\psi^*-i\ga\psi^*(|\psi|^2)_t,
  \end{split}
\end{equation}
so the perturbation terms are given by the expressions
\begin{equation}\label{eq22}
  R_{\psi}=i\ga\rho_t\psi,\qquad R_{\psi}=-i\ga\rho_t\psi^*.
\end{equation}
$A_m$ is the highest order of the derivative in the expression for $\mathcal{A}$, that is
in case of Eq.~(\ref{eq10}) we have $A_m=2$.
The Whitham velocities $v_i$ of unperturbed equations are equal to
\begin{equation}\label{eq23}
  \frac1{v_i}=-\frac{\langle \mathcal{B}/{{g}}\rangle}{\langle1/{{g}}\rangle},\qquad
  i=1,\ldots,4.
\end{equation}
In this expressions and in the right-hand side of Eq.~(\ref{eq20}) we have to put $\la=\la_i$.
Then we easily get
\begin{equation}\label{eq24}
  \Big\langle\frac1{ \tilde{g}}\Big\rangle_{\la=\la_i}=-\frac{2}{T}\frac{\prt T}{\prt \la_i}
\end{equation}
and
\begin{equation}\label{eq25}
  \frac1{v_i}=\frac1V-\frac{T}{2\prt_iT},\quad \prt_i\equiv\frac{\prt}{\prt\la_i},
  \quad i=1,2,3,4,
\end{equation}
or in the explicit form
\begin{equation} \label{eq26}
\begin{split}
    & \frac{1}{v_1}= \frac{1}{2}\sum_{i=1}^{4}\lambda_i-\frac{(\lambda_4-\lambda_1)
    (\lambda_2-\lambda_1)K}{(\lambda_4-\lambda_1)K-(\lambda_4-\lambda_2)E}, \\
    & \frac{1}{v_2}= \frac{1}{2}\sum_{i=1}^{4}\lambda_i+\frac{(\lambda_3-\lambda_2)
    (\lambda_2-\lambda_1)K}{(\lambda_3-\lambda_2)K-(\lambda_3-\lambda_1)E}, \\
    & \frac{1}{v_3}= \frac{1}{2}\sum_{i=1}^{4}\lambda_i-\frac{(\lambda_4-\lambda_3)
    (\lambda_3-\lambda_2)K}{(\lambda_3-\lambda_2)K-(\lambda_4-\lambda_2)E}, \\
    & \frac{1}{v_4}= \frac{1}{2}\sum_{i=1}^{4}\lambda_i+\frac{(\lambda_4-\lambda_3)
    (\lambda_4-\lambda_1)K}{(\lambda_4-\lambda_1)K-(\lambda_3-\lambda_1)E},
\end{split}
\end{equation}
where $E=E(m)$ is the complete elliptic integral of the second kind.

To calculate the right-hand side of Eq.~(\ref{eq20}), we substitute Eq.~(\ref{eq10})
for $\mathcal{A}$ and (\ref{eq22}) for $R_m$ and obtain after evident simplifications
the following formula for the expression in curly brackets
\begin{equation}\label{eq27}
\begin{split}
  &\frac{\prt\mathcal{A}}{\prt\psi}R_{\psi}+
  \frac{\prt\mathcal{A}}{\prt\psi_t}\frac{\prt R_{\psi}}{\prt t}+
  \frac{\prt\mathcal{A}}{\prt\psi_{tt}}\frac{\prt^2R_{\psi}}{\prt t^2}+
  \frac{\prt\mathcal{A}}{\prt\psi^*}R_{\psi^*}=\\
  &=\la\rho_{tt}+\frac{i\psi_t}{2\psi}\rho_{tt}-\frac{i}{2}\rho_{ttt}.
  \end{split}
\end{equation}
Now we can substitute Eq.~(\ref{eq19}) and equations $\rho_t=2\sqrt{\mathcal{R}}$,
$\rho_{tt}=2\mathcal{R}'(\rho)$, $\rho_{ttt}=4\mathcal{R}^{\prime\prime}(\rho)\sqrt{\mathcal{R}}$
in order to express the right-hand side as a function of $\rho$. At last, the averaging has to
be done according to the rule
\begin{equation}\label{eq28}
  \langle \mathcal{F}\rangle=\frac1T\int_0^T \mathrm{}\mathcal{F}dt
  =\frac1T\oint \mathcal{F}(\rho)\frac{d\rho}{2\sqrt{\mathcal{R}(\rho)}},
\end{equation}
where $\rho$ goes around a contour that encircles the segment $\nu_1\leq\rho\leq\nu_2$ in the
complex $\rho$-plane. As a result, we arrive at a lengthy expression which we write down here as a
sum of three integrals
\begin{equation}\label{eq29}
  \langle\dots\rangle=\frac{1}{T}\left(I_1+I_2+I_3\right),
\end{equation}
where
\begin{equation}\label{eq30}
  \begin{split}
   I_1=&\la\oint\mathcal{R}'\left(\la-\frac{s_1}{4}+\frac{j}{2\rho}-\frac{i}{2\rho}\sqrt{\mathcal{R}}\right)
  \frac{d\rho}{\sqrt{\mathcal{R}}},\\
   I_2=&\oint\mathcal{R}'\left(-\frac{s_1}{4}-\frac{j}{2\rho}+\frac{i}{2\rho}\sqrt{\mathcal{R}}\right)\\
  &\times\left(\la-\frac{s_1}{4}+\frac{j}{2\rho}-\frac{i}{2\rho}\sqrt{\mathcal{R}}\right)\frac{d\rho}{\sqrt{\mathcal{R}}},\\
  I_3=&-i\oint\mathcal{R}^{\prime\prime}\left(\la-\frac{s_1}{4}+\frac{j}{2\rho}-\frac{i}{2\rho}\sqrt{\mathcal{R}}\right)d\rho.
  \end{split}
\end{equation}
In transformations of these expressions, we take into account that integrals over closed contours of
single-valued functions and of full differentials vanish. As a result, after evident simplifications,
we obtain
\begin{equation}\label{eq31}
\begin{split}
  I_1+I_2+I_3=&-\frac13\oint \frac{\mathcal{R}^{3/2}d\rho}{\rho^3}-j^2\oint\frac{\sqrt{\mathcal{R}}d\rho}{\rho^3}\\
  &+\frac14\oint\frac{\mathcal{R}^{\prime2}d\rho}{\rho\sqrt{\mathcal{R}}}.
  \end{split}
\end{equation}
The integrals here can be expressed in terms of complete elliptic ones, but in practice it is easier
to deal with them in not integrated form as $\oint\mathcal{F}d\rho=2\int_{\nu_1}^{\nu_2}\mathcal{F}d\rho$.
As a result, we arrive at the Whitham modulation equations in the form
\begin{equation}\label{eq32}
\begin{split}
  \frac{\prt\la_i}{\prt x}+\left(\frac{s_1}{2}-\frac{T}{2\prt_iT}\right)\frac{\prt\la_i}{\prt t}=
  -\frac{T}{2\prt_iT}\frac{\ga}{\prod_{k\neq i}(\la_i-\la_k)}\frac{Q}{T},
  \end{split}
\end{equation}
where
\begin{equation}\label{eq33}
\begin{split}
  Q=&-\frac23\int_{\nu_1}^{\nu_2} \frac{\mathcal{R}^{3/2}d\rho}{\rho^3}-
  2j^2\int_{\nu_1}^{\nu_2}\frac{\sqrt{\mathcal{R}}d\rho}{\rho^3}\\
  &+\frac12\int_{\nu_1}^{\nu_2}\frac{\mathcal{R}^{\prime2}d\rho}{\rho\sqrt{\mathcal{R}}}.
  \end{split}
\end{equation}

Now we can turn to finding their stationary solution that describes a dispersive shock.

\section{A stationary DSW}\label{sec4}

A stationary shock propagates with constant velocity, so that the parameters $\la_i$
have the form of a traveling wave, $\la_i=\la_i(\xi)$, $\xi=t-x/V_1$. In fact, it is
similar to the well-known viscous shock (see, e.g., Refs.~\cite{whitham-74,kamch-2024}), but
now a sharp transition region of strong dissipation is replaced by a lengthy region of
modulated oscillations whose profile is determined by the Whitham equations.

It is essential that the factor $Q$ in Eqs.~(\ref{eq32}) is the same for all $i=1,2,3,4$.
Consequently, as it is easy to see, Eqs.~(\ref{eq32}) have a solution with $1/V_1=1/V=s_1/2$,
$\xi=\theta$, provided $\la_i=\la_i(\theta)$ satisfy the equations
\begin{equation}\label{eq34}
  \frac{d\la_i}{d\theta}=\frac{\ga}{\prod_{k\neq i}(\la_i-\la_k)}\frac{Q}{T},\quad i=1,2,3,4,
\end{equation}
and $s_1=\mathrm{const}$ is an integral of these equations. To prove this, we use the identity
\begin{equation}\label{eq35}
  \sum_{i=1}^4\frac{\prod_{k=1}^4(\la-\la_k)}{\prod_{k=1}^4(\la_i-\la_k)}=1,
\end{equation}
which is obviously correct, since the left-hand side is a polynomial in $\la$ of 3rd degree
equal to unity at four points $\la=\la_i$ and, consequently, equal to unity identically.
Comparing coefficients of $\la^n$ at both sides of this identity, we obtain after evident
simplifications the identities
\begin{equation}\label{eq36}
  \begin{split}
  & \sum_{i=1}^4\frac{1}{\prod_{k=1}^4(\la_i-\la_k)}=0,\\
  & \sum_{i=1}^4\frac{\la_i}{\prod_{k=1}^4(\la_i-\la_k)}=0,\\
  & \sum_{i=1}^4\frac{\la_i^2}{\prod_{k=1}^4(\la_i-\la_k)}=0,
  \end{split}
\end{equation}
and
\begin{equation}\label{eq37}
  \sum_{i=1}^4\frac{\la_1\la_2\la_3\la_4}{\la_i\prod_{k=1}^4(\la_i-\la_k)}=-1.
\end{equation}
The identities (\ref{eq36}) yield at once
\begin{equation}\label{eq38}
  \frac{d\sigma_n}{d\theta}=0\quad \text{for}\quad \sigma_n=\sum_{i=1}^{4}\la_i^n,\quad n=1,2,3,
\end{equation}
and since $\sigma_n$ are related to $s_n$ by the Newton formulas
\begin{equation}\label{eq39}
\begin{split}
  & s_1=\sigma_1,\quad s_2=\frac12(\sigma_1^2-\sigma_2),\\
  & s_3=\frac16(\sigma_1^3-3\sigma_1\sigma_2+2\sigma_3),
  \end{split}
\end{equation}
we obtain three integrals of the system (\ref{eq34}),
\begin{equation}\label{eq40}
  s_1=\mathrm{const},\quad s_2=\mathrm{const},\quad s_3=\mathrm{const}.
\end{equation}
A simple calculation gives the equations for $s_4=\la_1\la_2\la_3\la_4$,
\begin{equation}\label{eq41}
  \frac{ds_4}{d\theta}=-\frac{\ga Q}{T}.
\end{equation}
Thus, the stationary DSW is described by a single ordinary differential equation (\ref{eq41}),
where $\la_i=\la_i(s_4)$ are functions of $s_4$ defined as roots of the 4th degree algebraic
equation
\begin{equation}\label{eq42}
  P(\lambda)=\lambda^4-s_1\lambda^3+s_2\lambda^2-s_3\lambda+s_4=0
\end{equation}
with constant coefficients $s_1,s_2,s_3$. The values of these constant coefficients are
determined by the boundary conditions for the shock.

Generally speaking, an initial discontinuity evolves to a combination of two waves, and each can
be either a rarefaction wave or a DSW (see, e.g., \cite{LL-6,kamch-2024}). These two waves are
joined by a plateau whose parameters are determined by special conditions, which are different
for rarefaction waves, viscous shocks, and DSWs. A rarefaction wave is a simple wave solution
of dispersionless (hydrodynamic) equations, and this means that one of the Riemann  invariants
(\ref{eq4d}) preserves its value across such a wave, and this gives one of the boundary conditions
for the plateau parameters. In the standard theory of viscous shocks, we have the well-known
Rankine-Hugoniot jump conditions across a shock, and this gives another boundary condition for
the plateau parameters. These two conditions yield full classification of possible structures
that can evolve from an initial discontinuity in the theory of viscous shocks (see, e.g.,
\cite{LL-6,kamch-2024}). The situation is different in the case of DSWs. In the Gurevich-Pitaevskii
theory of DSWs for completely integrable equations, only one Riemann invariant of the Whitham
equations changes along a DSW in a self-similar solution of the Whitham equations, and this
provides a necessary second boundary condition, so we arrive at full classifications of appearing
wave structures, including generalizations on not genuinely nonlinear equations (see, e.g.,
\cite{eggk-95,kkl-12,ikcp-17,kamch-18}). However, when we take into account a weak dissipation,
the DSW solution is not self-similar anymore, and all Riemann invariants are changing along it.
On the other hand, instead of jump conditions, now we have several integrals of the Whitham
equations, and these integrals replace the Rankine-Hugoniot conditions (see, e.g., \cite{egk-05}).
In our case, these integrals are given by Eqs.~(\ref{eq40}). It is worth noticing that Gurevich and
Meshcherkin supposed in Ref.~\cite{gm-84}, on the basis of numerical experiments, that even in the
case of not completely integrable equations, the value of one of the dispersionless Riemann
invariants is transferred through a DSW in spite of the fact that Whitham equations cannot be
transformed to diagonal Riemann form. This supposition was used in Ref.~\cite{ik-19} for the
analytical description of DSWs induced by the Raman effect, and agreement with numerical
simulations was quite satisfactory. We develop here a more consistent theory based on the
preservation of all three integrals (\ref{eq40}) as the replacement of the Rankine-Hugoniot
conditions of the usual theory of viscous shocks.

\begin{figure}[t] \centering
\includegraphics[width=8cm]{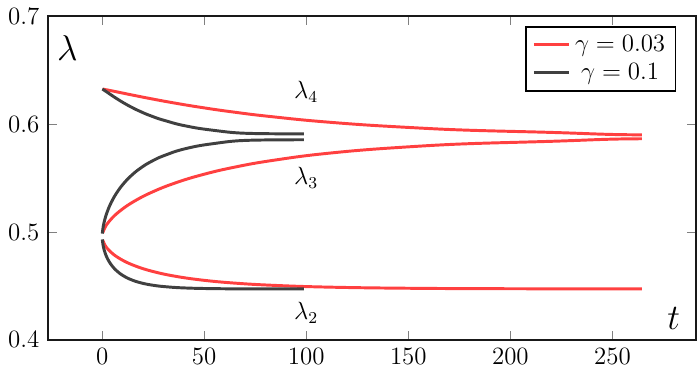}
\caption{Dependence of the Riemann invariants $\la_i,i=2,3,4$ on $t$ for two values of
the coefficient $\ga=0.03$ and $\ga=0.1$. The invariant $\la_1$ is practically constant
($\la_1\cong-0.447$) and it is not shown here.}
\label{fig1}
\end{figure}

As we shall see, the left edge $t\to-\infty$ of a stationary DSW corresponds to the trailing soliton,
so at this edge we have the matching conditions
\begin{equation}\label{eq43}
  \la_2^L=\la_3^L=z,\quad \la_1^L=\la_-^L,\quad \la_4^L=\la_+^L.
\end{equation}
The right edge of the DSW tends to a uniform state as $t\to+\infty$, so we get here other
matching conditions
\begin{equation}\label{eq44}
  \la_3=\la_4=y,\quad \la_1^R=\la_-^R,\quad \la_2^R=\la_+^R,
\end{equation}
where $z$ and $y$ are still unknown. The integrals (\ref{eq40}) give the relationships
\begin{equation}\label{eq45}
\begin{split}
&2z+\la_-^L+\la_+^L = 2y+\la_-^R+\la_+^R,\\
&z^2+2(\la_-^L+\la_+^L)z+\la_-^L\la_+^L = y^2+2(\la_-^R+\la_+^R)y+\la_-^R\la_+^R.
\end{split}
\end{equation}
and
\begin{equation}\label{eq46}
  z^2\la_-^L\la_+^L = y^2\la_-^R\la_+^R.
\end{equation}
Eqs.~(\ref{eq45}) yield the expressions
\begin{equation}\label{eq47}
  \begin{split}
  z= &\frac{(\la_-^L-\la_+^L)^2-(\la_-^R+\la_+^R)^2}{2(\la_-^L+\la_+^L-\la_-^R-\la_+^R)}\\
  &-\frac{(\la_-^R+\la_+^R)(\la_-^L+\la_+^L)-(\la_-^R)^2-(\la_+^R)^2}{\la_-^L+\la_+^L-\la_-^R-\la_+^R},\\
   y= &\frac{(\la_-^L+\la_+^L)^2-(\la_-^R-\la_+^R)^2}{2(\la_-^L+\la_+^L-\la_-^R-\la_+^R)}\\
  &+\frac{(\la_-^R+\la_+^R)(\la_-^L+\la_+^L)-(\la_-^L)^2-(\la_+^L)^2}{\la_-^L+\la_+^L-\la_-^R-\la_+^R}
  \end{split}
\end{equation}
for $z$ and $y$. Their substitution into Eq.~(\ref{eq46}) gives the relation between
the dispersionless parameters $\la_-^L,\la_+^L,\la_-^R,\la_+^R$ at both sides of the shock.
This relationship plays exactly the same role as the Rankine-Hugoniot condition in the theory
of standard viscous shocks---a single DSW connects two states with a certain relationship
between the flow parameters at its sides.

\begin{figure}[t] \centering
\includegraphics[width=8cm]{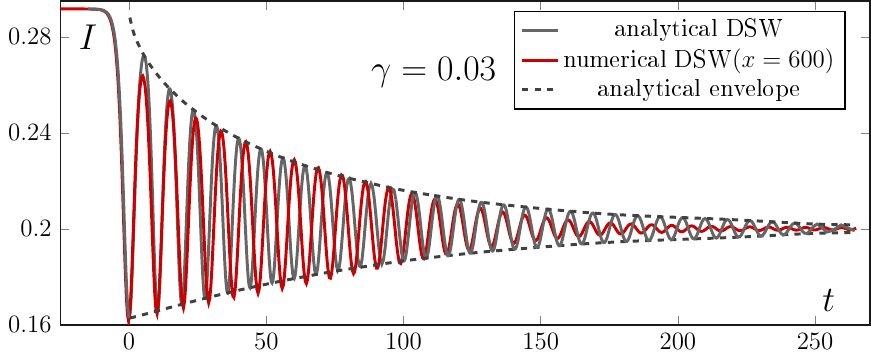}
\caption{Profile of intensity along the DSW for $\ga=0.03$.
}
\label{fig2}
\end{figure}

\begin{figure}[t] \centering
\includegraphics[width=8cm]{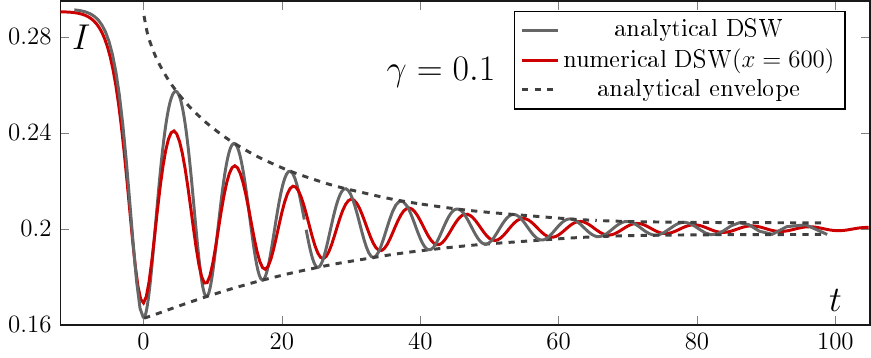}
\caption{Profile of intensity along the DSW for $\ga=0.1$.}
\label{fig3}
\end{figure}

Thus, in order to find distributions of the wave variables in a stationary DSW,
we choose the boundary conditions which satisfy Eqs.~(\ref{eq46}) and (\ref{eq47}),
so that the values of integrals $s_1,s_2,s_3$ are also known. Then we find $\la_i,i=1,2,3,4,$ 
as functions $\la_i=\la_i(s_4)$ of $s_4$ from Eq.~(\ref{eq42}), and, consequently,
the right-hand side of Eq.~(\ref{eq41}) becomes a known function of $s_4$. As a result,
we can solve Eq.~(\ref{eq41}) numerically and find the solution $s_4=s_4(\theta)$.
Therefore, the functions $\la_i=\la_i[s_4(\theta)]$ are also known and their substitution
into Eqs.~(\ref{eq5}) yields the profiles $\rho=\rho(\theta)$ and $u=u(\theta)$. In particular,
as is clear from Eq.~(\ref{eq8b}), the amplitude of the trailing soliton is equal to
\begin{equation}\label{eq48}
  a=(\la_+^L-z)(z-\la_-^L).
\end{equation}

\begin{figure}[t] \centering
\includegraphics[width=8cm]{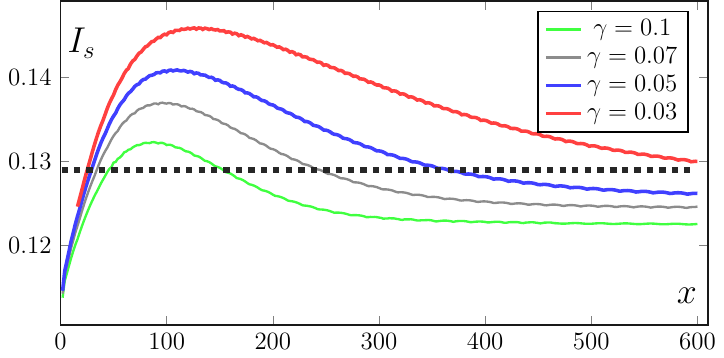}
\caption{Dependence of the trailing soliton amplitude (\ref{eq48}) on the propagation distance
for different values of $\ga$. The dashed thick line shows the theoretical value calculated in
the limit of very small $\ga$.}
\label{fig4}
\end{figure}

To compare our theory with exact numerical solution of Eq.~(\ref{eq1}), we have chosen the
parameters $\la_+^L=\sqrt{0.4}, \la_{\pm}^R=\pm\sqrt{0.2}$, so that they satisfy the
conditions (\ref{eq46}), (\ref{eq47}). The resulting dependence of  $\la_i,i=2,3,4,$ on $\theta$
is shown in Fig.~\ref{fig1}. The parameter $\la_1\cong-0.447$ remains practically constant along 
this DSW, so the supposition of Ref.~\cite{gm-84} is approximately fulfilled. Distribution of
intensity $\rho$ for $\ga=0.03$ is shown in Fig.~\ref{fig2} and for $\ga=0.1$ in Fig.~\ref{fig3}.
As one can see, agreement of our analytical theory with exact numerical solution is quite
good almost everywhere except the vicinity of the left edge, where the Whitham averaging method
looses its accuracy at distances of the order of magnitude of one wavelength. At last, we show
in Fig.~\ref{fig4}, how the amplitude of the trailing soliton (\ref{eq48}) changes with distance
of DSW's propagation for different values of $\ga$. As one can see, the curves come nearer to the
theoretical value $a\approx 0.13$ shown by a dashed line in the limit of very small $\ga$, as it
should be in our perturbation approach.

\section{Conclusion}

In recent years, several experiments have been performed in optical or similar systems specially
designed for demonstration of DSWs (see, e.g., \cite{xckt-17,mossman-18,bienaime-21,Bendahmane-22}).
In general, these experimental results agree quite well with earlier theoretical predictions,
provided the conditions of their applicability were fulfilled. For example, in Ref.~\cite{mossman-18},
Bose-Einstein condensate was confined in a quasi-1D trap, but the axial confinement was not strong
enough to exclude axial dynamics, so decay of shocks to vortices was effective and the shock had a
standard viscous form rather than that of a DSW. In Ref.~\cite{bienaime-21} dissipation was also
strong enough, but DSWs were clearly seen, although their profiles could only be calculated numerically.
In fibers, the one-dimensional geometry is evident and dissipative effects are negligibly small,
so the general structure reproduces theoretical predictions perfectly well \cite{xckt-17,Bendahmane-22}.
In this case, some weaker effects can become crucially important for long-distance propagation of pulses.
In the physics of optical fibers, the most important such effects are self-steepening and Raman
scattering (see, e.g., Ref.~\cite{ka-03}). As was shown in Ref.~\cite{ik-17}, the steepening effect
changes parameters of the DSW, but it preserves the expanding evolution of the shock. On the contrary,
the Raman effect can stabilize such an expansion, so the shock acquires a stationary form of a
modulated oscillatory profile moving with constant velocity. In the small-amplitude limit, the
theory of such shocks is described by the KdV-Burgers equation \cite{kivshar-90,km-93}, so the
results of Refs.~\cite{gp-87,akn-87,mg-95,kamch-16} can be applied. In this paper, we have developed
the theory of DSWs induced by the Raman effect for large amplitudes. The Whitham equations are derived
and thoroughly studied. It is shown that they have enough number of conservation laws for finding the
parameters of the shock for given boundary conditions that have to satisfy the analogue of the Rankine-Hugoniot
condition. It is important that this Rankine-Hugoniot condition is not universal in the sense that it
does not follow from conservation laws for the hydrodynamic equations, as it happens in the classical
theory of viscous shocks. In our case, the hydrodynamic equations have the form of ``shallow water
equations'' (\ref{eq4b}), the same as in the case of the Kaup-Boussinesq-Burgers equations \cite{egk-05},
but the sets of conservation laws of the Whitham equations in these two cases are different, and,
consequently, the Rankine-Hugoniot conditions are different, too. Thus, the theory developed in this
paper yields both the method of finding the analogues of the Rankine-Hugoniot conditions for completely
integrable equations with dissipative perturbations and the method of calculation of stationary profiles
of wave variables. One may hope that this theory can find other interesting applications.

\section*{Acknowledgments}

This research was funded  by the RSF grant number~19-72-30028 (Sections~1-3).
AMK was also supported by the research project FFUU-2021-0003 of the Institute of Spectroscopy
of the Russian Academy of Sciences (Sections~4-5).


\begin{thebibliography}{99}

\bibitem{eh-16} G. A. El and M. A. Hoefer, Dispersive shock waves and modulation theory.
Physica D, {\bf 333,} 11 (2016).

\bibitem{kamch-21} A.~M.~Kamchatnov, Gurevich-Pitaevskii problem and its development,
Usp. Fiz. Nauk., {\bf 191,} 52-87 (2021) [Phys.--Uspekhi, {\bf 64,} 48-82 (2021)].

\bibitem{gp-73} A. V. Gurevich and L. P. Pitaevskii,  Nonstationary structure of a collisionless shock wave,
Zh. Eksp. Teor. Fiz., {\bf 65,} 590 (1973) [Sov. Phys.-JETP, \textbf{38}, 291 (1974)].

\bibitem{whitham-65} G. B. Whitham, Non-linear dispersive waves, Proc. Roy. Soc. London, A {\bf 283}, 238 (1965).

\bibitem{whitham-74} G.~B.~Whitham, {\it Linear and Nonlinear Waves,}
(Wiley Interscience, New York, 1974).

\bibitem{gle-90} A. V. Gurevich, A. L. Krylov, and G. A. El, Nonlinear modulated waves in dispersive hydrodynamics,
Zh. Eksp. Teor. Fiz. {\bf 98,} 1605-1626 (1990) [Sov. Phys. JETP, {\bf 71,} 899-910 (1990)].

\bibitem{wright-93} O. C. Wright, Korteweg-de Vries Zero Dispersion Limit: Through First Breaking for Cubic-Like
Analytic Initial Data, Commun. Pure Appl. Math., {\bf 46,} 423-440 (1993).

\bibitem{tian-93} F. R. Tian, Oscillations of the Zero Dispersion Limit of the Korteweg-de Vries Equation,
Commun. Pure Appl. Math., {\bf 46,} 1093-1129 (1993).

\bibitem{gp-87} A. V. Gurevich, L. P. Pitaevskii, Averaged description of waves in the
Korteweg-de Vries-Burgers equation, Zh. Eksp. Teor. Fiz. {\bf 93,} 871 (1987)
[Sov. Phys. JETP, {\bf 66,} 490 (1987)].

\bibitem{akn-87} V. V. Avilov, I. M. Krichever, S. P. Novikov, Evolution of Whitham's zone in
Kortewed-de Vries theory, Dokl. Akad. Nauk SSSR, {\bf 295,} 345-349 (1987)
[Sov. Phys. Dokl. {\bf 32,} 564 (1987)].

\bibitem{mg-95} S. Myint, R. Grimshaw, The modelation of nonlinear periodic wavetrains by
dissipative terms in the Korteweg-de Vries equation, Wave Motion, {\bf 22,} 215-238 (1995).

\bibitem{kamch-16} A. M. Kamchatnov, Whitham theory for perturbed Korteweg–de Vries equation,
Physica D,  {\bf 333,} 99-106 (2016).

\bibitem{tsj-85} W. J. Tomlinson, R. H. Stolen, and A. M. Johnson, Optical wave breaking of pulses
in nonlinear optical fibers,
Opt. Lett. {\bf 10,} 457 (1985).

\bibitem{rg-89} J. E. Rothenberg and D. Grischkowsky, Observation of the Formation of an Optical
Intensity Shock and Wave Breaking in the Nonlinear Propagation of Pnlses in Optical Fibers,
Phys. Rev. Lett. {\bf 62,} 531-534 (1989).

\bibitem{ggkm-67} C. S. Gardner, J. M. Green, M. D. Kruskal and R. M. Miura,
Method for solving the Korteweg-de Vries equation, Phys. Rev. Lett., {\bf 19,} 1095 (1967).

\bibitem{lax-68} P. D. Lax, Integrals of Nonlinear Equations of Evolution and Solitary Waves,
Commun. Pure Appl. Math., {\bf 21,} 467 (1968).

\bibitem{zs-71} V. E. Zakharov and A. B. Shabat, Exact theory of two-dimensional self-focusing and
one-dimensional self-modulation of waves in nonlinear media, Zh. Eksp. Teor. Fiz., {\bf 61,} 118 (1971);
[Sov. Phys. JETP, {\bf 34,} 62 (1972)].

\bibitem{zs-73} V. E. Zakharov and A. B. Shabat,  Interaction between solitons in a stable medium,
Zh. Eksp. Teor. Fiz., {\bf 64,} 1627 (1973); [Sov. Phys. JETP, {\bf 37,} 823 (1973)].

\bibitem{fl-86} M. G. Forest and J. E. Lee, Geometry and modulation theory for periodic nonlinear
Schr\"{o}dinger equation, in {\it Oscillation Theory, Computation,
and Methods of Compensated Compactness,} edited by C. Dafermos, J. L. Ericksen, D. Kinderlehrer, and M. Slemrod, IMA
Volumes on Mathematics and its Applications Vol. 2 (Springer, New York, 1986), p. 35.

\bibitem{pavlov-87} M. V. Pavlov, Nonlinear Schr\"{o}dinger equation and Bogoliubov-Whitham averaging
method,  Teor. Mat. Fiz. {\bf 71,} 351-356 (1987) [Theor. Math. Phys. {\bf 71,} 584 (1987)].

\bibitem{ak-87} A. V. Gurevich and A. L. Krylov, Dissipationless shock waves in media with positive dispersion,
Zh. Eksp. Teor. Fiz., {\bf 92,} 1684-1699 (1987); [Sov. Phys. JETP {\bf 65,} 944-953 (1987)].

\bibitem{eggk-95} G. A. El, V. V. Geogjaev, A. V. Gurevich, and A. L. Krylov, Decay of an initial discontinuity
in the defocusing NLS hydrodynamics, Physica D  {\bf 87,} 186-192 (1995).

\bibitem{xckt-17} G. Xu, M. Conforti, A. Kudlinski, A. Mussot, and S. Trillo, Dispersive Dam-Break Flow of a Photon Fluid,
Phys. Rev. Lett. {\bf 118,} 254101 (2017).

\bibitem{ikp-19} M. Isoard, A. M. Kamchatnov, and N. Pavloff, Wave breaking and formation of dispersive
shock waves in a defocusing nonlinear optical material,
Phys. Rev. A {\bf 99,} 053819 (2019).

\bibitem{bienaime-21} T. Bienaime, M. Isoard, Q. Fontaine, A. Bramati, A.M. Kamchatnov, Q. Glorieux, and N. Pavloff,
Quantitative Analysis of Shock Wave Dynamics in a Fluid of Light,
Phys. Rev. Lett. {\bf 126,} 183901 (2021).

\bibitem{lpk-12} P.-\'{E}. Larr\'{e}, N. Pavloff, and A. M. Kamchatnov, Wave pattern induced by a localized
obstacle in the flow of a one-dimensional polariton condensate,
Phys. Rev. B {\bf 86,} 165304 (2012).

\bibitem{dianov-85} E. M. Dianov, S. Ya. Karasik, P. V. Mamyshev, A. M. Prokhorov, V. N. Serkin, M. F. Stel'makh, A. A. Fomichev,
Stimulated-Raman conversion of multisoliton pulses in quartz optical fibers,
Pis'ma Zh. Eksp. Teor. Fiz., {\bf 41,} 242-244 (1985); [JETP Lett., {\bf 41,} 294-297 (1985)].

\bibitem{mm-86} F. M. Mitschke and L. F. Mollenauer, Discovery of the soliton self-frequency shift,
Opt. Lett., {\bf 11,} 659-661 (1986).

\bibitem{weiner-89} A. M. Weiner, R. N. Thurston, W. J. Tomlinson, J. P. Heritage, D. E. Leaird, E. M. Kirschner,
and R. J. Hawkins, Temporal and spectral self-shifts of dark optical solitons,
Opt. Lett., {\bf 14,} 868-870 (1989).

\bibitem{stolen-89} R. H. Stolen, J. P. Gordon, W. J. Tomlinson, and H. A. Haus,
Raman response function of silica-core fibers,
J. Opt. Soc. Am., B {\bf 6,} 1159-1166 (1989).

\bibitem{kivshar-90} Y. S. Kivshar, Dark-soliton dynamics and shock waves induced by the stimulated
Raman efFect in optical fibers, Phys. Rev. A {\bf 42,} 1757-1761 (1990).

\bibitem{ka-03} Y. S. Kivshar and G. P. Agrawal, {\it Optical Solitons. From fibers to Photonic Crystals,}
(Academic Press, Amsterdam, 2003).

\bibitem{km-93} Y. S. Kivshar and B. A. Malomed, Raman-induced optical shocks in nonlinear fibers,
Opt. Lett., {\bf 18,} 485-487 (1993).

\bibitem{kamch-2000} A.~M.~Kamchatnov, {\it Nonlinear Periodic Waves and Their Modulations---An
Introductory Course,} (World Scientific, Singapore, 2000).

\bibitem{kamch-2024} A.~M.~Kamchatnov, {\it Theory of Nonlinear Waves,} (Moscow, HSE, 2024) [in Russian].

\bibitem{kamch-04} A. M. Kamchatnov, On Whitham theory for perturbed integrable equations,
Physica D, {\bf 188,} 247 (2004).

\bibitem{akns-74} M. J. Ablowitz, D. J. Kaup, A. C. Newell, H. Segur,
The Inverse Scattering Transform---Fourier Analysis for Nonlinear Problems,
Stud. Appl. Math., {\bf 53,} 249 (1974).

\bibitem{egk-05} G. A. El, R. H. J. Grimshaw, A. M. Kamchatnov,
Analytic model for a weakly dissipative shallow-water undular bore,
Chaos, {\bf 15,} 037102 (2005).

\bibitem{LL-6} L. D. Landau and E. M. Lifshitz, {\it Fluid Mechanics,} (Pergamon,
Oxford, 1987).

\bibitem{kkl-12} A. M. Kamchatnov, Y.-H. Kuo, T.-C. Lin, T.-L. Horng, S.-C. Gou, R. Clift, G. A. El,
and R. H. J. Grimshaw, Undular bore theory for the Gardner equation,
Phys. Rev. E {\bf 86,}  036605 (2012).

\bibitem{ikcp-17} S. K. Ivanov, A. M. Kamchatnov, T. Congy, N. Pavloff, Solution of the Riemann problem
for polarization waves in a two-component Bose-Einstein condensate, Phys. Rev. E. {\bf 96,} 062201 (2017).

\bibitem{kamch-18} A. M. Kamchatnov, Evolution of initial discontinuities in the DNLS equation theory,
J. Phys. Commun. {\bf 2,} 025027 (2018).

\bibitem{gm-84} A. V. Gurevich and A. P. Meshcherkin, Expanding self-similar discontinuities and shock
waves in dispersive hydrodynamics,
Zh. Eksp. Teor. Fiz., {\bf 87,} 1277-1292 (1984) [Sov. Phys.-JETP, \textbf{60}, 732-740 (1984)].

\bibitem{ik-19} S. K. Ivanov and A. M. Kamchatnov, Evolution of photon fluid wave pulses with account
of Raman effect, Optics and Spectroscopy, {\bf 127,} 101-111 (2019).

\bibitem{mossman-18} M. E. Mossman, M. A. Hoefer, K. Julien, P. G. Kevrekidis, and P. Engels,
Dissipative Shock Waves Generated by a Quantum Mechanical Piston,
Nat. Commun. {\bf 9,} 4665 (2018).

\bibitem{Bendahmane-22} A. Bendahmane, G. Xu, M. Conforti, A. Kudlinski, A. Mussot, and S. Trillo,
The piston Riemann problem in a photon superfluid,
Nat. Commun. {\bf 13,} 3137 (2022).

\bibitem{ik-17} S. K. Ivanov, A. M. Kamchatnov, Riemann problem for the photon fluid: Self-steepening effects,
Phys. Rev. A {\bf 96,} 053844 (2017).


\end{thebibliography}
\end{document}